\documentclass[useAMS,usegraphicx]{mn2e}

\usepackage{multirow}
\usepackage{bigdelim}
\usepackage{longtable}
\usepackage{graphicx}
\usepackage{times}

\def \form13{H$_2^{13}$CO }

\def \formd2{D$_2$CO }

\def \met13{$^{13}$CH$_3$OH }

\def \metan2d{CHD$_2$OH }
\def \metan3d{CH$_3$OD }

\usepackage{fixltx2e}
\title[]{Nitrogen oxide in protostellar envelopes 
and shocks: the ASAI survey}

\author[C. Codella et al.]{C. Codella$^{1}$\thanks{E-mail:
codella@arcetri.astro.it}, S. Viti$^{2}$, B. Lefloch$^{3}$,
J. Holdship$^{2}$, R. Bachiller$^{4}$, E. Bianchi$^{1,5}$, 
C. Ceccarelli$^{3}$, \newauthor 
C. Favre$^{1}$, I. Jim\'enez-Serra$^{6}$, 
L. Podio$^{1}$, M. Tafalla$^{4}$  \\
\\
$^{1}$ INAF-Osservatorio Astrofisico di Arcetri, L.go E. Fermi 5, Firenze, 50125, Italy \\
$^{2}$ Department of Physics and Astronomy, University College London, Gower Street, London, WC1E 6BT, UK \\
$^{3}$ Univ. Grenoble Alpes, CNRS, Institut de
Plan\'etologie et d'Astrophysique de Grenoble (IPAG), 38000 Grenoble, France \\
$^{4}$ IGN, Observatorio Astron\'omico Nacional, Calle Alfonso XII, 28004 Madrid, Spain \\
$^{5}$ Dipartimento di Fisica e Astronomia, Universit\'a degli Studi di Firenze, Italy \\
$^{6}$ School of Physics and Astronomy, Queen Mary University of London, 327 Mile End Road, London, E1 4NS
}

\begin{document}

\date{Accepted date. Received date; in original form date}

\pagerange{\pageref{firstpage}--\pageref{lastpage}} \pubyear{2016}

\maketitle

\label{firstpage}

\begin{abstract}

The high-sensitivity of the IRAM 30-m ASAI unbiased spectral survey
in the mm-window allows
us to detect NO emission towards both the Class I object SVS13-A 
and the protostellar outflow shock L1157-B1.
We detect the hyperfine
components of the $^2\Pi_{\rm 1/2}$
$J$ = 3/2 $\to$ 1/2 (at 151 GHz)
and the $^2\Pi_{\rm 1/2}$ $J$ = 5/2 $\to$ 3/2 (250 GHz) spectral
pattern. The two objects show different NO profiles:
(i) SVS13-A emits through narrow (1.5 km s$^{-1}$) lines 
at the systemic velocity, while (ii) L1157-B1 shows 
broad ($\sim$ 5 km s$^{-1}$) blue-shifted emission.
For SVS13-A the analysis leads to  
$T_{\rm ex}$ $\geq$ 4 K, 
$N(\rm NO)$ $\leq$ 3 $\times$ 10$^{15}$ cm$^{-2}$,
and indicates the 
association of NO with the protostellar envelope.
In L1157-B1, NO is tracing the extended
outflow cavity: $T_{\rm ex}$ $\simeq$ 4--5 K, and 
$N(\rm NO)$ = 5.5$\pm$1.5 $\times$ 10$^{15}$ cm$^{-2}$.
Using C$^{18}$O, $^{13}$C$^{18}$O, C$^{17}$O, and $^{13}$C$^{17}$O
ASAI observations we derive 
an NO fractional abundance less than $\sim$ 10$^{-7}$
for the SVS13-A envelope, in agreement with previous measurements
towards extended PDRs and prestellar objects. 
Conversely, a definite $X(NO)$ enhancement is measured 
towards L1157-B1, $\sim$ 6 $\times$ 10$^{-6}$,
showing that the NO production increases in shocks.
The public code UCLCHEM 
was used to interpret the NO
observations, confirming that 
the abundance observed in SVS13-A can be attained in 
an envelope with
a gas density of 10$^5$ cm$^{-3}$ and a kinetic temperature of 40 K.
The NO abundance in L1157-B1
is reproduced with pre-shock densities of 10$^5$ cm$^{-3}$
subjected to a $\sim$ 45 km s$^{-1}$ shock.
\end{abstract}

\begin{keywords}
Stars: formation -- ISM: molecules -- ISM: abundances -- ISM: jets and outflows 
\end{keywords}

\section{Introduction}

The molecule of NO, Nitrogen Oxide, was first detected 
towards SgrB2 (Liszt \& Turner 1978) and the molecular cloud OMC1 
(Blake et al. 1986). Although it has now been detected in a variety 
of objects, ranging from dark clouds, to hot cores, photon-dominated 
regions and nuclei of starburst galaxies 
(e.g. Gerin et al. 1992, 1993; Halfen et al. 2001; Nummelin et al. 2000; 
Martin et al. 2003), it is seldom observed, possibly due to 
its complex hyperfine structure. Nevertheless, NO may be an important 
chemical tracer of the nitrogen chemistry cycle (e.g. Hily-Blant et al. 2010) 
as well as of the oxygen budget (Chen et al. 2014).

Although the chemistry of nitrogen-bearing species has been modelled in 
several environments (e.g. Flower et al. 2006; Hily-Blant et al. 2010; 
Awad et al. 2016), 
the total nitrogen abundance in the gas phase is still uncertain, especially 
in dense cores, as well as in the gas surrounding protostars.
This is because much of the nitrogen is in the form of N and/or N$_2$, 
both not directly observable. Generally NH, NH$_2$ and NH$_3$ are used as 
probes of the total nitrogen reservoir in dense star forming gas. 
However the abundance of these three species is highly dependent on how 
much nitrogen depletes on the surface of the grains and on the level of 
hydrogenation.
In fact in some environments NO seems to be one of the most abundant 
N-bearing species: for example, Velilla Prieto et al. (2015) finds that 
in circumstellar oxigen-rich envelopes NO is as high as 10$^{-6}$ and one of the most 
abundant N-bearing species. 

NO is key to form N$_2$, a molecule required to then form
more complex nitrogen-bearing species (Hily-Blant et al. 2010) 
and harbours a simpler chemistry than other N-bearing species
such as NH, NH$_2$ and NH$_3$. 
The most direct way to form NO is 
via neutral-neutral reactions between N and OH and, hence, 
its abundance mainly depends on the amount of available nitrogen in the gas.
Finally in environments dominated by low velocity shocks ($\sim$ 20 kms$^{-1}$) NO seems to also be a direct 
probe of the molecular oxygen abundance. Indeed, 
low velocity shock models indicate 
that the behaviour of NO is similar to that of O$_2$ in 
that they are abundant (and deficient) at similar times (Chen et al. 2014). 
Since O$_2$ is not readily observable with ground-based telescopes, 
NO may prove to be a good proxy for the molecular form of oxygen.

In this paper we present NO detections towards two objects: 
the SVS13-A Class I system and the 
outflow spot L1157-B1. Both environments are warm, but they 
significantly differ from their chemical composition and past history. 
These two regions are described in detail in Section 2. 
In Section 3 we describe the observations. In Section 4 we report the 
results of our excitation analysis. In Section 5 we derive the observed 
fractional abundances of NO and we model the chemistry of NO to find the 
origin of its emission in both objects. We briefly 
present our conclusions in Section 6.
 
\section{The sample}

\subsection{The Class I SVS13-A object}

SVS13-A is a $\simeq$ 32.5 $L_{\rm \odot}$ (Tobin et al. 2016)
young star located in a Young Stellar Objects (YSOs) cluster in the 
Perseus NGC1333 cloud ($d$ = 235 pc; Hirota et al. 2008). The region has been 
extensively investigated in the past
(see e.g. Chini et al. 1997; Bachiller et al. 1998; Looney et al.
2000; Chen et al. 2009; Tobin et al. 2016, and references therein).
Figure 1 reports recent 1.4mm continuum interferometric 
observations of SVS13 (De Simone et al. 2017) showing SVS13-A
as well as the SVS13-B protostar, which lies out of the region
sampled with the present IRAM 30-m observations (see Sect. 3). 
SVS13-A is associated
(i) with a chemically rich hot-corino 
(Codella et al. 2016; De Simone et al. 2017; Lef\`evre et al. 2017), 
(ii) with an extended outflow 
($>$ 0.07 pc, Lefloch et al. 1998, Codella et
al. 1999) as well as (iii) with the well-known chain of
Herbig-Haro (HH) objects 7-11 (Reipurth et al. 1993). In addition,
SVS13-A has a low $L_{\rm submm}$/$L_{\rm bol}$ ratio ($\sim$ 0.8\%)
and a high bolometric temperature ($T_{\rm bol}$ $\sim$ 188 K, Tobin
et al. 2016).
Although still deeply embedded in a large scale molecular envelope
(Lefloch et al. 1998), SVS13-A is considered a more evolved
system, already entered in the Class I evolutionary stage
(e.g. Chen et al. 2009). Finally, iwe note that VLA observations
(e.g. Anglada et al. 2000; Tobin et al. 2016, and references therein) 
showed that SVS13-A, once observed at sub-arcsec angular scale,
reveals in turn two components (VLA4A and VLA4B) separated by 300 mas, being
the hot-corino associated only with VLA4B (Lef\`evre et al. 2017). 

\subsection{The L1157-B1 protostellar shock}

The L1157-mm Class 0 protostar, located at 250 pc from Earth 
(Looney et al. 2007) with a $L_{\rm bol}$ $\simeq$ 3 $L_{\rm \odot}$
(Tobin et al. 2010), drives a chemically rich outflow   
(e.g. Bachiller et al. 2001), associated with molecular clumpy cavities
(Gueth et al. 1996; 1998), that are created by
episodic events in a precessing jet (Podio et al. 2015).  
The brightest shock front is called B1 and is associated with the
southern blueshifted outflow lobe (see Fig. 2). 
The B1 structure has been studied
in detail using both single dishes and interferometers, revealing 
a clumpy bow-like structure, built by several jet-cavity impacts in the last $\sim$ 1000 years, where the abundance of several molecules are enhanced  
due to a combination of sputtering and warm gas-phase
chemistry (e.g. Tafalla \& Bachiller 1995; Codella et al. 2009; 
Benedettini et al. 2013; Busquet 
et al. 2014; Codella et al. 2015, Lefloch et al. 2017, and references therein).

\begin{figure}
\begin{center}
\includegraphics[angle=0,width=7cm]{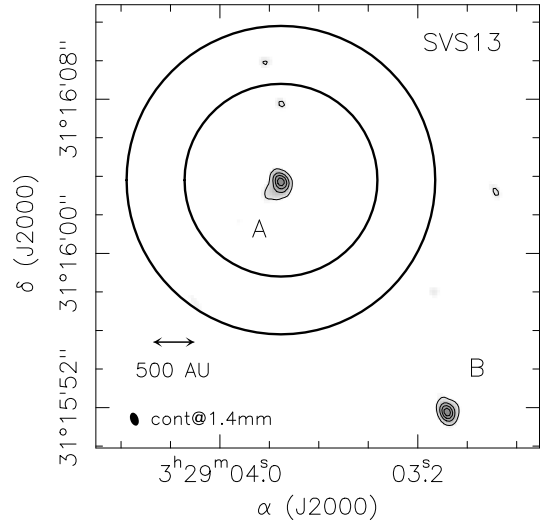}
\end{center}
\caption{The SVS13 star forming cluster as observed  
using continuum emission at 1.4mm with the PdBI by De Simone et al. (2017).
Two objects dominate at mm-wavelengths: the Class I system 
called A and the earlier Class 0 object called B.
The solid rings show the HPBWs of the IRAM 30-m for the
NO transitions detected at 2mm and 1mm towards SVS13-A.} 
\end{figure}

\begin{figure}
\begin{center}
\includegraphics[angle=0,width=7cm]{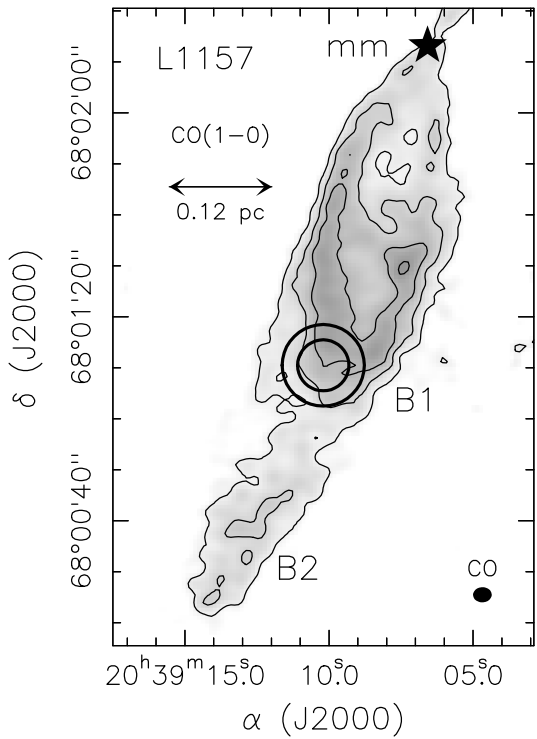}
\end{center}
\caption{The L1157 southern blueshifted outflow as traced by
CO(1--0) emission observed using the IRAM PdB interferometer
(Gueth et al. 1996). The B1 and B2 cavities are labelled.
The solid rings show the HPBWs of the IRAM 30-m for the
NO transitions detected at 2mm and 1mm towards B1 (see text).
The star marks the position of the L1157-mm protostar.}
\end{figure}

\section{Observations}

The observations of NO were carried out during several runs
between 2012 and 2014 with the IRAM 30-m telescope
near Pico Veleta (Spain) as part of 
the Astrochemical Surveys At IRAM\footnote{www.oan.es/asai} (ASAI)
Large Program. Both SVS13-A and L1157-B1 have been observed 
at 3 mm (80--116 GHz), 2 mm
(129--173 GHz), and 1.3 mm (200--276 GHz) using the  
broad-band EMIR receivers.
The observations were acquired towards the following coordinates:
SVS13-A: 
$\alpha_{\rm  J2000}$ = 03$^{\rm h}$ 29$^{\rm m}$ 03$\fs$3,
$\delta_{\rm J2000}$ = +31$\degr$ 16$\arcmin$ 03$\farcs$8;
L1157-B1:
$\alpha_{\rm  J2000}$ = 20$^{\rm h}$ 39$^{\rm m}$ 10$\fs$2,
$\delta_{\rm J2000}$ = +68$\degr$ 01$\arcmin$ 10$\farcs$5,
in wobbler switching mode, with a throw
of 180$\arcsec$.

The present study is based on the NO spectra of the $^2\Pi_{\rm 1/2}$ 
$J$ = 3/2 $\to$ 1/2 (at $\sim$ 150.2 GHz) and
$^2\Pi_{\rm 1/2}$ $J$ = 5/2 $\to$ 2/2 (at $\sim$ 250.8 GHz) 
transitions (see Table 1). 
The forward ($F_{\rm eff}$) and beam efficiencies ($B_{\rm eff}$) are:
0.93 and 0.72 at 150 GHz, and 0.90 and 0.53 at 251 GHz. 
We also report emission associated with the C$^{18}$O(2--1) (219560.350 MHz), 
C$^{17}$O(2--1)
(224714.190 MHz), $^{13}$C$^{18}$O(2--1) (209419.172 MHz),
$^{13}$C$^{17}$O(2--1) (214573.873 MHz), and CS(5--4) (244935.560 MHz) 
transitions detected towards SVS13-A. All the frequency values have
been extracted from the Cologne Database for Molecular Spectroscopy 
(CDMS; M\"uller et al. 2001, 2005) molecular database.
The pointing was checked by observing nearby planets or continuum
sources and was found to be accurate to within 2$\arcsec$--3$\arcsec$.
The telescope HPBWs are 16$\arcsec$ at 150.2 GHz and 10$\arcsec$
at 250.8 GHz. 
The data reduction was performed using the
GILDAS--CLASS\footnote{http://www.iram.fr/IRAMFR/GILDAS} package.
Calibration uncertainties are estimated to be $\simeq$ 20\%.  
Note that the lines observed at 2 mm towards SVS13-A 
(see Sect. 4) are affected by emission at OFF position observed in wobbler mode.
Line intensities have been converted from antenna temperature to main
beam temperature ($T_{\rm MB}$), using the main
beam efficiencies reported in the IRAM 30-m
website\footnote{http://www.iram.es/IRAMES/mainWiki/Iram30mEfficiencies}.

\section{Results}

\begin{table}
  \caption{List of NO transitions
detected$^a$ towards SVS13-A and L1157-B1 (see Table 2).}
  \begin{tabular}{lccccc}
  \hline
\multicolumn{2}{c}{Transition} &
\multicolumn{1}{c}{$\nu$$^{\rm b}$} &
\multicolumn{1}{c}{I$^{b,c}$} &
\multicolumn{2}{c}{Source}  \\ 
\multicolumn{1}{c}{$F_{\rm u}$ $\to$ $F_{\rm l}$} &
\multicolumn{1}{c}{$p_{\rm u}$ $\to$ $p_{\rm l}$} &
\multicolumn{1}{c}{(GHz)} &
\multicolumn{1}{c}{} &
\multicolumn{1}{c}{{SVS13-A}} &
\multicolumn{1}{c}{{L1157-B1}} \\ 
\hline
\multicolumn{6}{c}{NO $^2\Pi_{\rm 1/2}$ $J$ = 5/2 -- 3/2 ($E_{\rm u}$ = 19 K)} \\
\hline
7/2 $\to$ 5/2 & + $\to$ -- & 250436.85 & 99.7 & Y & Y \\
5/2 $\to$ 3/2 & + $\to$ -- & 250440.66 & 63.1 & Y & Y \\
3/2 $\to$ 1/2 & + $\to$ -- & 250448.53 & 37.4 & Y & Y \\
5/2 $\to$ 5/2 & -- $\to$ + & 250708.25 & 12.0 & N & Y \\
7/2 $\to$ 5/2 & -- $\to$ + & 250796.44 & 100.0 & Y & Y \\
5/2 $\to$ 3/2 & -- $\to$ + & 250815.59 & 63.3 & Y & Y \\
3/2 $\to$ 1/2 & -- $\to$ + & 250816.95 & 37.6 & Y & Y \\
\hline
\multicolumn{6}{c}{NO $^2\Pi_{\rm 1/2}$ $J$ = 3/2 -- 1/2 ($E_{\rm u}$ = 7 K)} \\
\hline
5/2 $\to$ 3/2 & -- $\to$ + & 150176.48 & 100.0 & Y & Y \\
3/2 $\to$ 1/2 & -- $\to$ + & 150198.76 & 37.1 & N & Y \\
3/2 $\to$ 3/2 & -- $\to$ + & 150218.73 & 29.6 & N & Y \\
1/2 $\to$ 1/2 & -- $\to$ + & 150225.66 & 29.6 & N & Y \\
3/2 $\to$ 3/2 & + $\to$ -- & 150439.12 & 29.6 & N & Y \\
5/2 $\to$ 3/2 & + $\to$ -- & 150546.52 & 100.0 & Y & Y \\
1/2 $\to$ 1/2 & + $\to$ -- & 150580.56 & 63.3 & N & Y \\
3/2 $\to$ 1/2 & + $\to$ -- & 150644.34 & 37.6 & N & Y \\
\hline
\end{tabular}

$^a$ The detected lines have a S/N ratio $\ge$ 5 and are
not contaminated by the emission from other lines (see text).
$^b$ From the Cologne Database
 for Molecular Spectroscopy 
(CDMS; http://www.astro.uni-koeln.de/cdms/;
 M\"uller et al. 2001, 2005) molecular database.
$^c$ $I$ is the expected intensity. We assumed $I$ = 100.0 for the brightest
component; the intensities of the other hyperfine components are
consequently scaled.
\end{table}

The NO emission due to $^2\Pi_{\rm 1/2}$
$J$ = 3/2 $\to$ 1/2 and $^2\Pi_{\rm 1/2}$ $J$ = 5/2 $\to$ 3/2 
transitions has been detected towards both the
SVS13-A and L1157-B1 sources. In particular, Table 1 reports 
all the hyperfine components detected above 5$\sigma$. 
Figures 3, 4, 5, and 6 reports the observed spectra, 
that are discussed below for both sources.

We used the GILDAS CLASS tool to fit separately the NO 
$^2\Pi_{\rm 1/2}$
$J$ = 3/2 $\to$ 1/2 and $^2\Pi_{\rm 1/2}$ $J$ = 5/2 $\to$ 3/2 spectra, 
obtaining 
the best fit of the 
hyperfine components providing four parameters
(see Codella et al. 2012): (i) the LSR velocity,
(ii) the linewidth (FWHM), (iii) the sum of the opacity at the central
velocities of all the hyperfine
components $p_1=\sum_{\rm i}$$\tau_{\rm i}$,
and (iv) the product $p_2$ = $p_1$ $\times$ [$J$($T_{\rm ex}$)--$J$($T_{\rm bg}$)--$J$($T_{\rm c}$)],
where $J(T) = \frac{h\nu/k}{{\rm e}^{h\nu/kT}-1}$ and $T_{\rm c}$ 
is the temperature of the continuum emission, which can be neglected for
both sources here analysed (see below).
Hence:

\begin{equation}
T_{ex} = \frac{h\nu}{k} \left[ln \left(1+\frac{h\nu}{k}\frac{p_1}{p_2}\right)\right]^{-1}.
\end{equation}

The derived $T_{\rm ex}$ depends also on the
assumed source extent to reproduce the observed
$T_{\rm mb}$ and will be discussed separately for each source.

\subsection{SVS13-A}

Figure 3 shows in black the emission lines due to the
NO $^2\Pi_{\rm 1/2}$ $J$ = 5/2 -- 3/2 hyperfine components
observed at $\simeq$ 250 GHz (Table 1). The HPBW of 10$\arcsec$
ensures no contamination due to the SVS13-B protostar (see Fig. 1).  
The red line shows the fit obtained with the CLASS tool:
the lines are well peaked at the systemic velocity of +8.6
km s$^{-1}$ (Chen et al. 2009), the FWHM is quite narrow
(1.5$\pm$0.1 km s$^{-1}$), and 
the emission looks optically thin.
Note that the $F$ = 5/2 -- 3/2 and
3/2 -- 1/2 (-- $\to$ +) transitions of the $^2\Pi_{\rm 1/2}$ $J$ = 5/2 -- 3/2
pattern are blended at
the present observed resolution, as clearly shown in Fig. 3.

Which is the size of the region emitting in NO?
Instructive information comes from the linewidth, which is
indeed narrower than that tracing the inner 100 AU
of the protostellar environment, i.e. the hot-corino,
which is 4--5 km s$^{-1}$ accordingly to recent
single-dish and interferometric observations of SVS13-A
(L\'opez-Sepulcre et al. 2015; Codella et al. 2016; 
De Simone et al. 2017).
On the other hand, the NO linewidth (1.5 km s$^{-1}$) is in perfect
agreement with that of CS(5--4), a tracer of the molecular
envelope (see Table 3 and Fig. 7). Indeed, Lefloch et al. (1998)
mapped the SVS13-A region in CS(5--4) using the IRAM 30-m antenna
revealing an envelope with size of 
$\simeq$ 39$\arcsec$ $\times$ 17$\arcsec$ 
($\simeq$ 0.04 $\times$ 0.02 pc). 
The temperature has been estimated from
continuum to be around 40 K
(Lefloch et al. 1998; Chen et al. 2009).
In addition, also the C$^{18}$O and C$^{17}$O lines
here detected in the context of ASAI  
(Table 3 and Fig. 7) also show 
similar FWHM.
All these findings support the association of 
the NO $^2\Pi_{\rm 1/2}$ $J$ = 5/2 -- 3/2 emission with
the molecular envelope surrounding the SVS13-A protostar.
Consequently, we applied no filling factor correction
since the envelope size is larger than 17$\arcsec$.
We then derive the excitation temperature $T_{\rm ex}$ by
considering the opacity less than 0.2. 
The continuum
emission, roughly estimated from the high quality baselines
obtained by the Wobbler mode observations ($\sim$ 80 mK in $T_{MB}$ scale),
can be neglected.
We obtain low $T_{\rm ex}$, $\geq$ 4 K, 
clearly indicating that the LTE conditions
are not fully satisified and that the density of the 
observed envelope
has to be less than 
the critical densities of the observed NO lines, which are,
according to the collisional rates with H$_2$ derived
by Lique et al. (2009),
$\sim$ 10$^5$ cm$^{-3}$ in the 10--50 K range
(see http://home.strw.leidenuniv.nl/~moldata/datafiles/no.dat). 
In addition, it has to be noted that the low-excitation
($E_{\rm u}$ $<$ 20 K) does not make the observed NO lines 
the best candidates to probe high kinetic temperatures.
The total column density of NO is then estimated to be
$N(\rm NO)$ $\leq$ 3 $\times$ 10$^{15}$ cm$^{-2}$.

Interestingly enough, Figure 8 (bottom and middle panels) shows that if we
inspect the best Gaussian fit of the NO $^2\Pi_{\rm 1/2}$ $J$ = 5/2 -- 3/2
$F_{\rm ul}$ = 7/2 -- 5/2; 5/2 -- 3/2, and 3/2 -- 1/2; 
$p_{\rm ul}$ =  + $\to$ -- emission (i.e. the more isolated hyperfine
components of the $J$ = 5/2 -- 3/2 pattern), we see a
non-negligible residual at $\sim$ 3$\sigma$ (see the purple filled
histograms). These profiles suggest
non-Gaussian emission from gas at blue-shifted velocity.
Indeed the residual of the C$^{18}$O(2--1) and C$^{17}$O(2--1)
shows wings (Fig. 8-Upper) associated with the well known extended molecular 
blue- and red-shifted outflow driven by SVS13-A (e.g. Lefloch et al. 1998). 
The present NO $^2\Pi_{\rm 1/2}$ $J$ = 5/2 -- 3/2 emission 
thus suggests an outflow component, to be confirmed by higher sensitivity 
observations. 
If we conservatively assume an excitation temperature between 5 K and 50 K
and optically thin emission, we derive a column density  
$N(\rm NO)$ $\sim$ 10$^{14}$--10$^{15}$ cm$^{-2}$ in the wings.

Finally, in addition to the NO emission at $\sim$ 250 GHz, 
Fig.4 reports the spectral signature of the  
NO $^2\Pi_{\rm 1/2}$ $J$ = 3/2 -- 1/2 spectral pattern
observed towards SVS13-A with a HPBW = 16$\arcsec$.
Only the hyperfine component $F_{\rm ul}$ = 5/2 -- 3/2;
$p_{\rm ul}$ =  + $\to$ -- (150546.52 MHz), and 
$F_{\rm ul}$ = 5/2 -- 3/2;
$p_{\rm ul}$ =  -- $\to$ + (150176.48 MHz) are detected.
Both profiles are characterised by a weak emission centred at the
systemic velocity of +8.6 km s$^{-1}$, and are severely contaminated by a 
deep absorption due to emission in the OFF position, observed in wobbler mode 
with a throw of 180$\arcsec$ (see Sect. 3).
The absorption is present in the 150 GHz spectrum, taken
with a HPBW of 16$\arcsec$, and not at 251 GHz, where the beam 
is definitely smaller, i.e 10$\arcsec$.
However,
considered that the 150 GHz and 251 GHz observations have
been not observed simultaneously, different locations
of the OFF position around the ON source could have played a role.  
The absorption is blue-shifted suggesting contamination due
to gas associated the molecular
cloud hosting nearby protostellar systems (e.g.
the systemic velocity of the close NGC1333-IRAS4A and -IRAS2A objects 
is $\sim$ +7.2 km s$^{-1}$;
e.g. Santangelo et al. 2015; De Simone et al. 2017). 
The poor NO $^2\Pi_{\rm 1/2}$ $J$ = 3/2 -- 1/2 spectra prevent us 
to obtain a reliable fit. However, we verified that 
what found at $\sim$ 150 GHz is consistent with the solution provided by
the fit of the $^2\Pi_{\rm 1/2}$ $J$ = 5/2 -- 3/2:
$N(\rm NO)$  $\leq$ 3 $\times$ 10$^{15}$ cm$^{-2}$.

\begin{figure*}
\begin{center}
\includegraphics[angle=0,width=14cm]{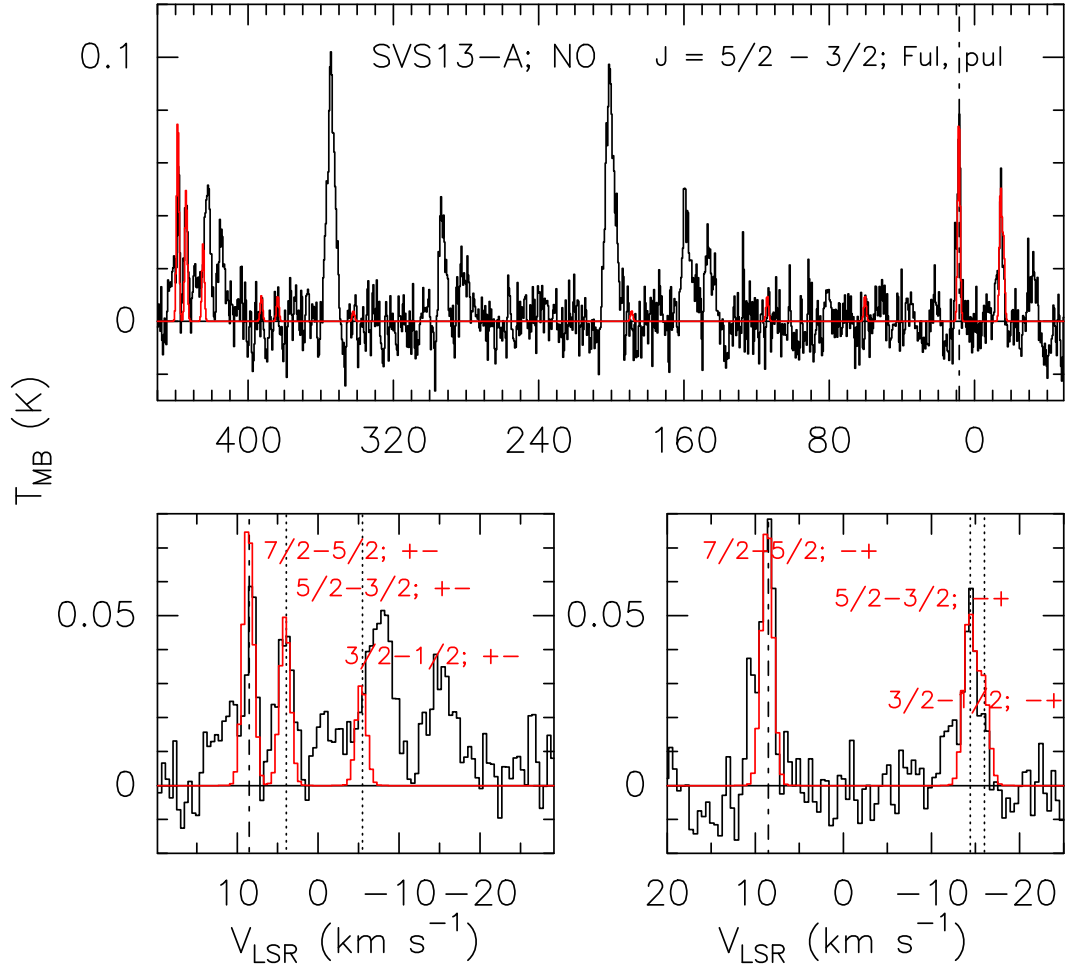}
\end{center}
\caption{{\it Upper panel:} Emission due to the 
NO $^2\Pi_{\rm 1/2}$ $J$ = 5/2 -- 3/2 spectral pattern  
(black histogram; see Table 1; in T$_{\rm mb}$ scale after
correction for beam dilution, see text) observed towards
SVS13-A with the IRAM 30-m antenna. 
The red line shows the best fit obtained with the 
GILDAS--Class package using the simultaneous 
fit of all the hyperfine components (see Table 2). 
For sake of clarity, the hyperfine components are not
indicated in the upper panel are labelled in the zoom-in in the lower
panels.
The spectrum is centred at the frequency of the
hyperfine component $F_{\rm ul}$ = 7/2 -- 5/2; 
$p_{\rm ul}$ =  + $\to$ -- (250436.85 MHz). 
The vertical dashed line
indicates the ambient LSR velocity (+8.6 km s$^{-1}$, Chen et al. 2009).
{\it Bottom-right panel:} Zoom in of the Upper panel. Dotted lines
indicate the hyperfine components falling in that portion of the spectrum.
Note that the $F_{\rm ul}$ = 3/2 -- 1/2
$p_{\rm ul}$ =  -- $\to$ + line is blended with the
brighter $F_{\rm ul}$ = 5/2 -- 1/2
$p_{\rm ul}$ =  -- $\to$ + one. The analysis of the residual is
shown in Fig. 8. 
{\it Bottom-left panel:} Zoom in of the Upper panel, after having centred 
the spectrum at the frequency of the
hyperfine component $F_{\rm ul}$ = 7/2 -- 5/2;  
$p_{\rm ul}$ =  + $\to$ -- (250796.44 MHz). 
The $F_{\rm ul}$ = 3/2 -- 1/2
$p_{\rm ul}$ =  + $\to$ -- line is blended with the 
13$_{\rm 9,4}$--12$_{\rm 9,3}$ intense emission at 250450.21 MHz due to CH$_3$CHO-E,
a well known tracer of hot-corinos (as that around SVS13-A).
For the legibility of the figure, given the complexity of this 
portion of the spectrum, the residual
is not reported.}
\end{figure*}

\begin{figure}
\begin{center}
\includegraphics[angle=0,width=6cm]{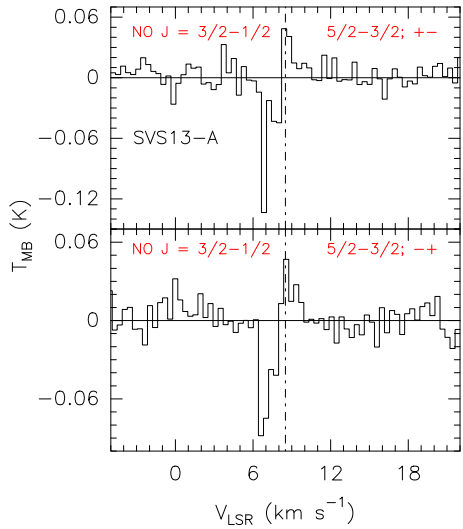}
\end{center}
\caption{{\it Upper panel:} Spectral signatures due to the
NO $^2\Pi_{\rm 1/2}$ $J$ = 3/2 -- 1/2 spectral pattern
(black histogram; see Table 1; in T$_{\rm mb}$ scale after
correction for beam dilution, see text) observed towards 
SVS13-A with the IRAM 30-m antenna.
The spectrum is centred at the frequency of the
hyperfine component $F_{\rm ul}$ = 5/2 -- 3/2;
$p_{\rm ul}$ =  + $\to$ -- (150546.52 MHz).
The vertical dashed line
indicates the ambient LSR velocity (+8.6 km s$^{-1}$, Chen et al. 2009).
The absorption is due to emission in the
OFF position, observed in wobbler mode 
with a throw of 180$\arcsec$ (see Sect. 3).
{\it Bottom panel:} Same as for the Upper panel for the
hyperfine component $F_{\rm ul}$ = 5/2 -- 3/2;
$p_{\rm ul}$ =  -- $\to$ + (150176.48 MHz).}
\end{figure}

\begin{table}
\caption{Parameters of the hyperfine fits to the 
observed NO lines (see Table 1).} 
\label{lines}
\centering
\begin{tabular}{cccccc}
\hline
\multicolumn{1}{c}{$T_{\rm ex}$} &
\multicolumn{1}{c}{rms$^a$} &
\multicolumn{1}{c}{$V_{\rm peak}$} &
\multicolumn{1}{c}{$FWHM$} &
\multicolumn{1}{c}{$\sum_{\rm i}$$\tau_{\rm i}$} &
\multicolumn{1}{c}{$N_{\rm tot}$$^b$} \\
\multicolumn{1}{c}{(K)} &
\multicolumn{1}{c}{(mK)} &
\multicolumn{1}{c}{(km s$^{-1}$)} &
\multicolumn{1}{c}{(km s$^{-1}$)} &
\multicolumn{1}{c}{} &
\multicolumn{1}{c}{(cm$^{-2}$)} \\
\hline
\hline
\multicolumn{6}{c}{SVS13-A} \\
\hline
\multicolumn{6}{c}{NO $^2\Pi_{\rm 1/2}$ $J$ = 5/2 -- 3/2} \\
\hline
$\geq$ 4  & 8 & +8.6$\pm$0.1) & 1.5$\pm$0.1 & $\leq$ 0.2 & $\leq$ 3 $\times$ 10$^{15}$ \\
\hline
\hline
\multicolumn{6}{c}{L1157-B1} \\
\hline
\multicolumn{6}{c}{NO $^2\Pi_{\rm 1/2}$ $J$ = 5/2 -- 3/2} \\
\hline
5$\pm$1  & 7 & +0.6$\pm$0.1 & 5.0$\pm$0.2 & 0.6$\pm$0.2 & 5.5$\pm$1.5 $\times$ 10$^{15}$ \\
\hline
\multicolumn{6}{c}{NO $^2\Pi_{\rm 1/2}$ $J$ = 3/2 -- 1/2} \\
\hline
$\geq$ 4  & 9 & +0.6$\pm$0.3 & 4.7$\pm$0.2 & $\leq$ 0.2 & $\leq$ 4 $\times$ 10$^{15}$ \\
\hline
\end{tabular}
\begin{center}
$^a$ At a spectral resolution of 0.47 km s$^{-1}$ (SVS13-A),
0.93 km s$^{-1}$ (L1157-B1; $^2\Pi_{\rm 1/2}$ $J$ = 5/2 -- 3/2),
and 0.78 km s$^{-1}$ (L1157-B1; $^2\Pi_{\rm 1/2}$ $J$ = 3/2 -- 1/2).
$^b$ Assuming Local Thermodynamic Equilibrium (LTE) and a source size of 
26$\arcsec$ (SVS13-A) and 20$\arcsec$ (L1157-B1; see text). \\
\end{center}
\end{table}

\begin{table*}
\caption{List of CO and CS transitions and 
line properties (in $T_{\rm MB}$ scale) detected towards SVS13-A and used in the present analysis.}
\begin{tabular}{lccccccccc}
\hline
\multicolumn{1}{c}{Transition} &
\multicolumn{1}{c}{$\nu$$^{\rm a}$} &
\multicolumn{1}{c}{$E_{\rm u}$} &
\multicolumn{1}{c}{$HPBW$} &
\multicolumn{1}{c}{$Dv$} &
\multicolumn{1}{c}{rms} &
\multicolumn{1}{c}{$T_{\rm peak}$$^b$} &
\multicolumn{1}{c}{$V_{\rm peak}$$^b$} &
\multicolumn{1}{c}{$FWHM$$^b$} &
\multicolumn{1}{c}{$I_{\rm int}$$^b$} \\
\multicolumn{1}{c}{ } &
\multicolumn{1}{c}{(MHz)} &
\multicolumn{1}{c}{(K)} &
\multicolumn{1}{c}{($\arcsec$)} &
\multicolumn{1}{c}{(km s$^{-1}$)} &
\multicolumn{1}{c}{(mK)} &
\multicolumn{1}{c}{(mK)} &
\multicolumn{1}{c}{(km s$^{-1}$)} &
\multicolumn{1}{c}{(km s$^{-1}$)} &
\multicolumn{1}{c}{(mK km s$^{-1}$)} \\
\hline
$^{13}$C$^{18}$O(2--1) & 209419.172 & 15 & 12 & 0.56 & 9 & 129$\pm$13 & +8.4$\pm$0.1 & 1.5$\pm$0.2 & 78$\pm$1 \\ 
$^{13}$C$^{17}$O(2--1) & 214573.873 & 15 & 12 & 0.55 & 10 & 18$\pm$3 & +8.2$\pm$0.4 & 1.7$\pm$0.9 & 32$\pm$15 \\
C$^{18}$O(2--1) & 219560.350 & 16 & 11 & 0.53 & 53 & 5370$\pm$150 & +8.6$\pm$0.1 & 1.6$\pm$0.1 & 8873$\pm$74 \\
C$^{17}$O(2--1) & 224714.190 & 16 & 11 & 0.52 & 18 & 1068$\pm$69 & +8.2$\pm$0.1 & 2.0$\pm$0.1 & 2315$\pm$28 \\ 
CS(5--4)        & 244935.560 & 35 & 10 & 0.49 & 20 & 386$\pm$71 & +8.3$\pm$0.1 & 1.4$\pm$0.1 & 592$\pm$26 \\
\hline
\end{tabular}

$^a$ From the Cologne Database
 for Molecular Spectroscopy
(CDMS; http://www.astro.uni-koeln.de/cdms/).
 M\"uller et al. 2001, 2005) molecular database. \\
$^b$ The errors are the Gaussian fit uncertainties (1$\sigma$) and are
in units of the last digit.. \\
\end{table*}

\begin{figure*}
\begin{center}
\includegraphics[angle=0,width=14cm]{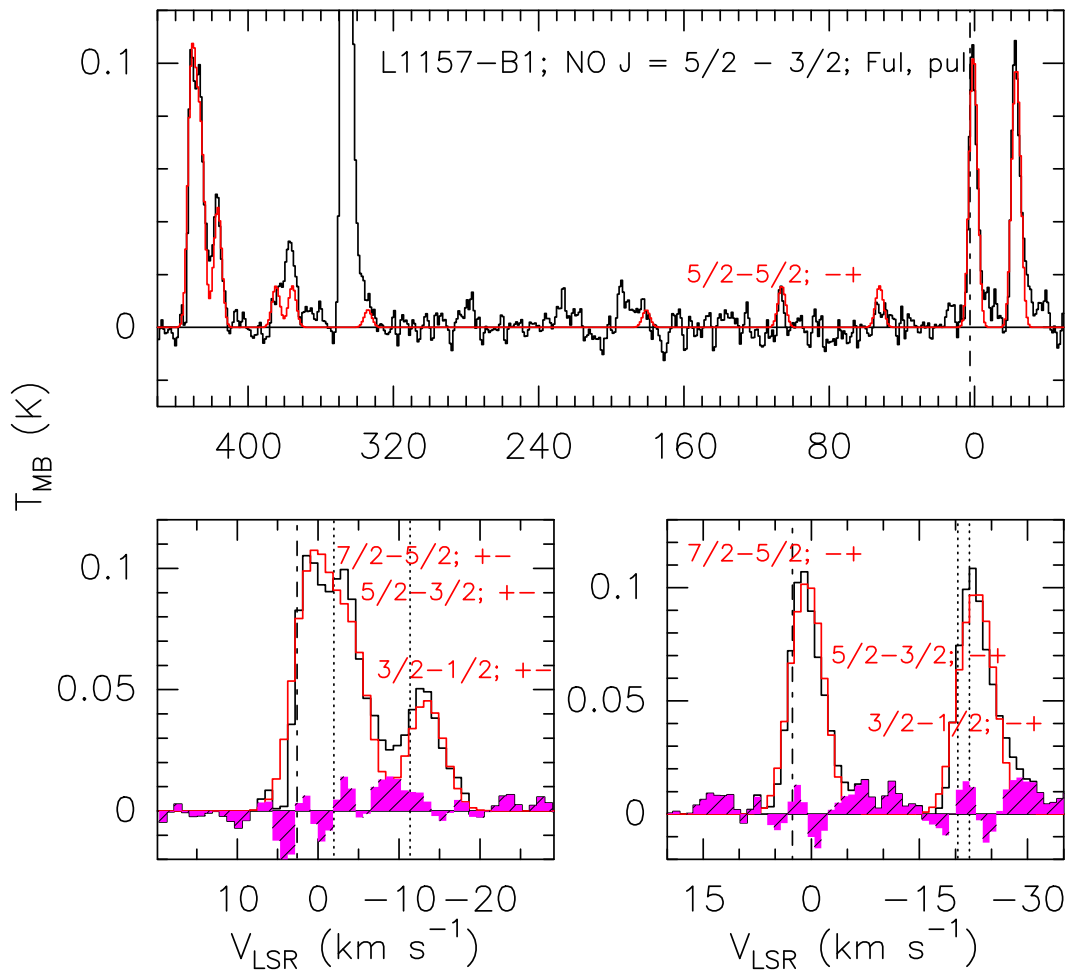}
\end{center}
\caption{{\it Upper panel:} Emission due to the
NO $^2\Pi_{\rm 1/2}$ $J$ = 5/2 -- 3/2 spectral pattern
(black histogram; see Table 1; in T$_{\rm mb}$ scale 
observed towards 
L1157-B1 with the IRAM 30-m antenna.
The red line shows the best fit obtained with the          
GILDAS--Class package using the simultaneous                    
fit of all the hyperfine components (see Table 2).
For sake of clarity, the hyperfine components which are not
indicated in the upper panel are labelled in the zoom-in in the lower
panels.
The spectrum is centred at the frequency of the
hyperfine component $F_{\rm ul}$ = 7/2 -- 5/2;
$p_{\rm ul}$ =  + $\to$ -- (250436.85 MHz).
The vertical dashed line
indicates the ambient LSR velocity (+2.6 km s$^{-1}$, Bachiller et al. 2001).
{\it Bottom-right panel:} Zoom in of the Upper panel. Dotted lines
indicate the hyperfine components falling in that portion of the spectrum.
Note that the $F_{\rm ul}$ = 3/2 -- 1/2 
$p_{\rm ul}$ =  -- $\to$ + line is blended with the
brighter $F_{\rm ul}$ = 5/2 -- 1/2 
$p_{\rm ul}$ =  -- $\to$ + one. The purple filled histogram shows the residual
after Gaussian fit (red) of the NO emission (black).
{\it Bottom-left panel:} Zoom in of the Upper panel, after having centred
the spectrum at the frequency of the
hyperfine component $F_{\rm ul}$ = 7/2 -- 5/2;
$p_{\rm ul}$ =  + $\to$ -- (250796.44 MHz).}
\end{figure*}

\begin{figure*}
\begin{center}
\includegraphics[angle=0,width=14cm]{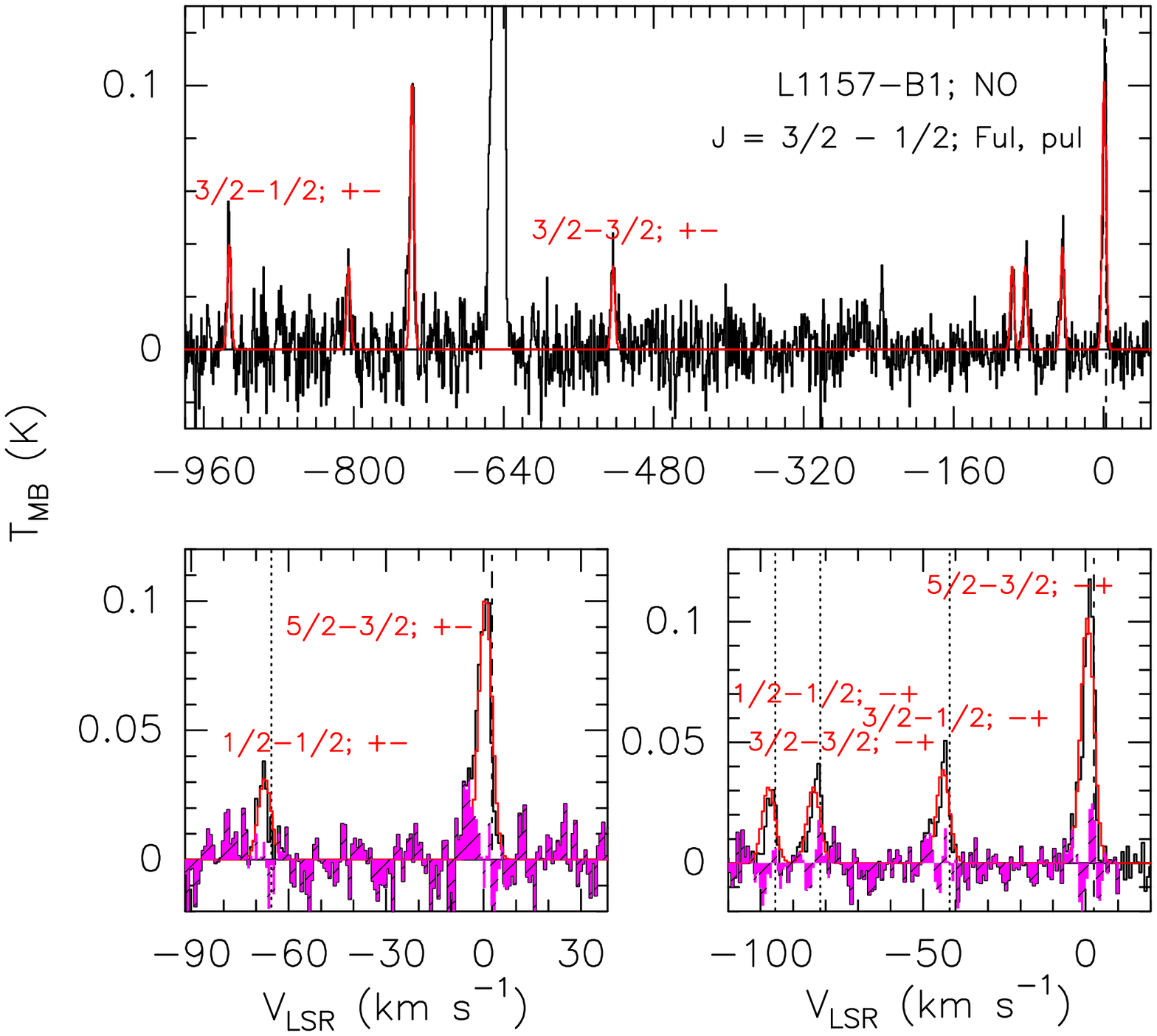}
\end{center}
\caption{{\it Upper panel:} Emission due to the
NO $^2\Pi_{\rm 1/2}$ $J$ = 3/2 -- 1/2 spectral pattern
(black histogram; see Table 1; in T$_{\rm mb}$ scale after
correction for beam dilution, see text) observed towards 
L1157-B1 with the IRAM 30-m antenna.
The red line shows the best fit obtained with the          
GILDAS--Class package using the simultaneous                    
fit of all the hyperfine components (see Table 2).
For sake of clarity, the hyperfine components which are not
indicated in the upper panel are labelled in the zoom-in in the lower
panels.
The spectrum is centred at the frequency of the
hyperfine component $F_{\rm ul}$ = 5/2 -- 3/2;
$p_{\rm ul}$ =  -- $\to$ + (150176.48 MHz).
The vertical dashed line
indicates the ambient LSR velocity (+2.6 km s$^{-1}$, Bachiller et al. 2001).
{\it Bottom-right panel:} Zoom in of the Upper panel. Dotted lines
points the hyperfine components falling in that portion of the spectrum.
The purple filled histogram shows the residual
after Gaussian fit (red) of the NO emission (black).
{\it Bottom-left panel:} Zoom in of the Upper panel, after having centred
the spectrum at the frequency of the
hyperfine component $F_{\rm ul}$ = 5/2 -- 3/2;
$p_{\rm ul}$ =  + $\to$ -- (150546.52 MHz).}
\end{figure*}

\begin{figure}
\begin{center}
\includegraphics[angle=0,width=6cm]{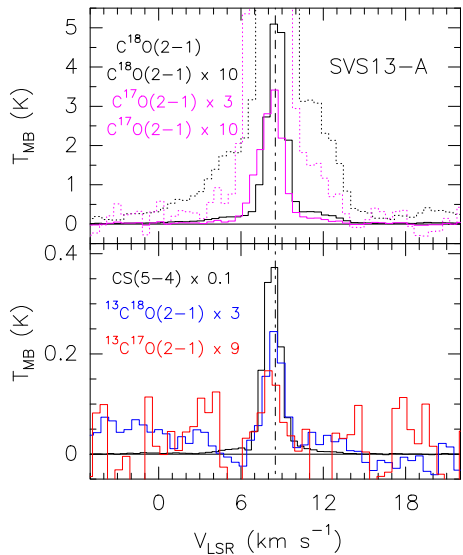}
\end{center}
\caption{Emission detected towards SVS13-A with the IRAM 30-m in the 1mm spectral window with
HPBWs consistent with that used for the
observations of NO $^2\Pi_{\rm 1/2}$ $J$ = 5/2 -- 3/2.
{\it Upper panel:} C$^{18}$O(2--1) and C$^{17}$O(2--1) lines
(in T$_{\rm mb}$ scale; also increased by different factors to show the outflow wings;
dotted histograms). The vertical dashed line
indicates the ambient LSR velocity (+8.6 km s$^{-1}$, Chen et al. 2009).
{\it Bottom panel:} $^{13}$C$^{18}$O(2--1), $^{13}$C$^{17}$O(2--1),
and CS(5--4) profiles, scaled to be drawn on the same scale.}
\end{figure}

\begin{figure}
\begin{center}
\includegraphics[angle=0,width=6cm]{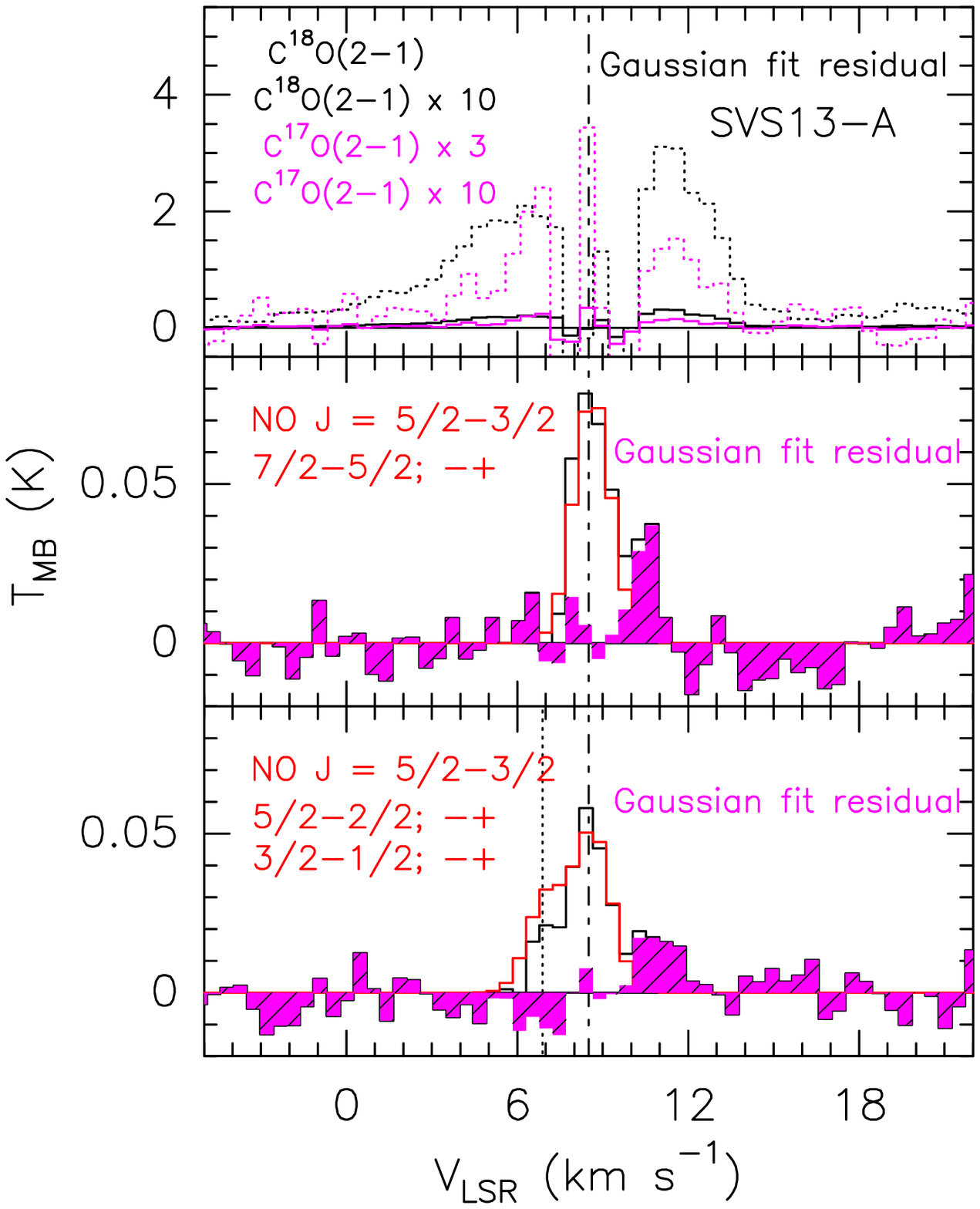}
\end{center}
\caption{{\it Upper panel:} Residual of the C$^{18}$O(2--1) and C$^{17}$O(2--1)
emission after Gaussian fit, showing the outflow wings. The vertical dashed line
indicates the ambient LSR velocity (+8.6 km s$^{-1}$, Chen et al. 2009).
{\it Middle and bottom panel:} The purple filled histogram shows the residual
after Gaussian fit (red) of the NO $^2\Pi_{\rm 1/2}$ $J$ = 5/2 -- 3/2
$F_{\rm ul}$ = 7/2 -- 5/2; 5/2 -- 3/2, and 3/2 -- 1/2; $p_{\rm ul}$ =  + $\to$ --
emission (black).}
\end{figure}

\subsection{L1157-B1}

Figures 5 and 6 show in black the emission lines due to the
NO $^2\Pi_{\rm 1/2}$ $J$ = 5/2 -- 3/2 and $J$ = 3/2 -- 1/2
hyperfine components, reported in Table 1.
The red lines show the fit of the NO hyperfine structure 
obtained with GILDAS-CLASS:
the emission lines of both spectral patterns have FWHMs $\simeq$ 5 km s$^{-1}$ and
peak at $\sim$ +0.6 km s$^{-1}$, being blue-shifted by about 2 km s$^{-1}$
($V_{\rm LSR}$ = +2.6 km s$^{-1}$; Bachiller et al. 2001).
Note that the brightest lines show blue wings 
(in addition to the Gaussian profile; see also the residuals
in Figs. 5 and 6) with velocities 
up to $\sim$ 8 km s$^{-1}$
from the systemic velocity.
As for SVS13-A, the $F$ = 5/2 -- 3/2 and
3/2 -- 1/2 (-- $\to$ +) transitions of the $^2\Pi_{\rm 1/2}$ $J$ = 5/2 -- 3/2
pattern are blended at
the present observed resolution (see Fig. 5). 
We thus report for the first time the association of NO emission
with a shocked region driven by a low-mass protostar.
Lefloch et al. (2012) showed that several components of L1157-B1
can be disentangled through the analysis of molecular
line profiles, namely: (i) the compact ($\geq$ 10$\arcsec$) gas
associated with the impact between the precessing jet and
the cavities, dominating at velocities less than --20 km s$^{-1}$;
(ii) the extended (20$\arcsec$) B1 cavity ($\simeq$ 1100 yr old; Podio
et al. 2016), dominating at velocities between 
about --20 km s$^{-1}$ and --5 km s$^{-1}$; (iii) the older 
extended (again $\sim$ 20$\arcsec$) cavity remnant associated
with the B2 jet episode, dominating at velocities close to the systemic one.  
The velocity range of the NO emission is in good 
agreement with that of several
molecular species (such as CS, H$_2$CO, NH$_3$; see
Bachiller et al. 2001; Codella et al. 2010; Benedettini et al. 2013;
G\'omez-Ruiz et al. 2015) tracing the extended B1 and B2 cavities,
outlined in Fig. 2 and encompassed by both the HPBWs of the present observations.
This will allow us to assume for the NO emission in L1157-B1
an emitting size of 20$\arcsec$ (the typical size of both the 
B1 and B2 emission)
and consequently to apply the filling factor correction.

To derive the excitation temperature $T_{\rm ex}$, the contribution due to
continuum can be neglected, given that no continuum emission has 
been so far detected towards L1157-B1 (e.g Codella et al. 2015, and
references therein). Once taken into account the opacities
derived from the Gaussian fit (opacity less than 0.6) we consistently derive for both the NO 
spectral patterns observed at $\sim$ 150 GHz and $\sim$ 250 GHz
an excitation temperature $T_{\rm ex}$ $\simeq$ 5 K. 
Also in this case, the low $T_{\rm ex}$ values suggest that 
the gas is not fully in LTE conditions. 
As reported before the critical densities of the observed NO lines  
at 250 GHz are $\sim$ 10$^5$ cm$^{-3}$ (Lique et al. 2009).
On the other hand, the critical densities of the 
transitions at 150 GHz are, in the 10--50 K range, 
are $\sim$ 10$^4$ cm$^{-3}$. 
These values are similar to the minimum
density derived by G\'omez-Ruiz et al. (2015) from CS for the extended
old ($\sim$ 2000 yr) cavity (5 $\times$ 10$^4$ cm$^{-3}$). 
Considered the measure of $T_{\rm ex}$ and the corresponding error, 
the total NO column density is $N(\rm NO)$ = 5.5$\pm$1.5 $\times$ 10$^{15}$ cm$^{-2}$.  

\section{Discussion}

\subsection{Abundances}

To estimate the NO abundance ($X_{\rm NO}$) in the
SVS13-A envelope, we evaluated the H$_2$ column
density using the emission from the $J$ = 2--1 transition of several
CO isotopologues (Table 3): C$^{18}$O, C$^{17}$O, and $^{13}$C$^{18}$O,
expected to be optically thin. We also detect
$^{13}$C$^{17}$O(2--1) emission with a Signal-to-Noise (S/N) ratio $\sim$ 3--4.
All these lines have been observed
with the same HPBWs (10$\arcsec$--12$\arcsec$) and show 
FWHMs consistent with that of the NO profiles (Table 3), indicating
that the rare CO isotopologues are reasonably tracing the same material,
i.e. the molecular envelope.
Adopting Local Thermodynamic Equilibrium (LTE) conditions,
optically thin emission, and assuming 
a kinetic temperature of 40 K (Lefloch et al. 1998), 
[$^{16}$O]/[$^{18}$C]=560, [$^{18}$O]/[$^{17}$C]=3, 
[$^{12}$C]/[$^{13}$C]=77, and
[$^{12}$CO]/[H$_2$] = 10$^{-4}$ (Wilson \& Rood 1994),
we derive $N({\rm CO})$ = 3--10 $\times$ 10$^{18}$ cm$^{-2}$.
Consequently, $X_{\rm NO}$ towards SVS13-A is   
$\leq$ 3 $\times$ 10$^{-7}$.
Interestingly, if we use the tentative NO detection towards
the SVS13-A low-velocity outflow and the C$^{17}$O emission
at the same velocities (see Fig. 8) we obtain
an NO abundance of 0.1--5 $\times$ 10$^{-7}$ i.e., a value 
equal or sligthly higher than that in the SVS13-A envelope. 

For the L1157-B1 analysis, we can use 
the source-averaged column density N(CO)
= 1 $\times$ 10$^{17}$ cm$^{-2}$, found for both the B1  
and B2 cavities by Lefloch et al. (2012). 
We can thus derive the NO abundance towards a shocked region: 
$X$(NO) = 4--7 $\times$ 10$^{-6}$. 
To our knowledge, there are few measurements of the NO abundances 
in low-mass star forming regions. Y{\i}ld{\i}z et al. (2013) reported
$X$(NO) = 2 $\times$ 10$^{-8}$ towards the NGC1333-IRAS4A star forming
region, suggesting NO is tracing an extended PDR region outside
the protostellar core. Similar abundances (1--3 $\times$ 10$^{-8}$)
have been reported for starless cores such as e.g. L1544
(Akylmaz et al. 2007).
The high abundance derived for L1157-B1 as well as that tentatively
derived for the SVS13-A outflow (fews 10$^{-7}$) 
supports a large increase of the NO production, either
due to direct release from dust mantles and/or by warm gas-phase
chemistry. 
In the following, we make use of the public code  
UCLCHEM (Holdship et al. 2017) to investigate 
the origin of the NO emission in each object. The code is fully described in Holdship et al. (2017) and references therein. Briefly, the code computes the fractional abundances of gas and solid species as a function of time. Beside the standard chemical gas phase reactions, freeze out, non thermal (due to UV radiation and cosmic rays), thermal desorption and a basic network of surface reactions are all included. 

\subsection{Chemical modelling of SVS13-A}

In the previous section we established that NO is almost certainly arising from  
the molecular envelope surrounding the SVS13-A protostar. In order to 
simulate the envelope, we run UCLCHEM in two Phases: 
in Phase 1 we start from a purely atomic gas at low density
(n$_H$ = 10$^2$ cm$^{-3}$), and allow the gas to collapse in
free-fall until a gas density of 10$^4$ is reached (the gas density estimated in Section 4.1).  The time it takes to reach this density is of course the free-fall timescale, which is $\sim$  5$\times$10$^6$ years. This Phase is then either stopped, or allowed to continue for a further 5 million years (assuming a typical lifetime for the parent molecular cloud of $\sim$ 10 Myr, Mouschovias et al. 2006).
We adopt the solar initial elemental abundances  
reported by Asplund et al. (2009), (see Table 4). 

\begin{table}
\caption{List of initial elemental fractional abundances (with respect to the total number of hydrogen nuclei) for Phase I of UCLCHEM, taken from Asplund et al. (2009)} 
\begin{tabular}{lc}
\hline
Species & Injected (/H) \\
\hline
He & 0.085 \\  
O & 4.898 $\times$ 10$^{-4}$ \\
C & 2.692 $\times$ 10$^{-4}$ \\
N & 6.761 $\times$ 10$^{-5}$ \\
S & 1.318 $\times$ 10$^{-5}$ \\
Mg & 3.981 $\times$ 10$^{-5}$ \\
Si & 3.236 $\times$ 10$^{-5}$ \\
Cl & 3.162 $\times$ 10$^{-7}$ \\
P & 2.57 $\times$ 10$^{-9}$ \\
F & 3.6 $\times$ 10$^{-8}$ \\
\hline
\end{tabular}
\end{table}

Phase 2, which starts at a density determined by the final step of Phase 1, then follows 
the chemical evolution as a function of time at constant density but at a temperature of 40 K,
as assumed in the 
previous sections and also in agreement with Lefloch et al. (1998). 
In Fig. 9 we show our results for Phase 2, following a Phase 1 of 10 Myr, for an envelope 
at 10$^4$ cm$^{-3}$:  NO steadily decreases with time and 
its abundance always remains below 10$^{-8}$ for times $>$ 2 $\times$ 10$^4$ years, in agreement with the derived 10$^{-7}$ upper limit. 
NO here forms via the neutral-neutral reaction of N + OH, and 
its behaviour indeed
follows closely that of OH which steadily decreases to form CO. 
We conclude that the observed NO can indeed be explained by the presence of an
envelope with a likely gas density of $\sim$ 10$^4$ cm$^{-3}$. 

Since we encompass the hot corino as well as 
the envelope of this protostar within our beam, it is important to verify 
whether a contribution from the hot corino should have been present. 
In order to model the hot corino we run UCLCHEM again in two phases; 
like before, in Phase 1 we start from a purely atomic gas at low density, 
and allow the gas to collapse in free-fall until a gas density of either 
10$^7$ cm$^{-3}$ or 10$^8$ cm$^{-3}$ is reached. 
In Phase 2 the dense gas is `placed' at a distance of $\sim$ 10 AU 
from the star and we allow the temperature to increase 
up to a value of 80--100~K. 
The increase in temperature with time is set up in a similar manner 
to the 1 M$_{\sun}$ model by Awad et al. (2010), but with the assumption that the final mass of the 
star is 0.75 M$_{\sun}$, as determined from the 1.4 mm dust continuum observations of Chen et al. (2009). 
As the temperature increases, the mantle sublimates 
in various temperature bands (see Collings et al. 2004). Phase 2 follows 
the gas and mantle chemical evolution with time  up to 10$^5$ years,  
a reasonable age of a protostar entering the Class I stage.
In order to exclude a contribution from the hot corino, 
the NO fractional abundance with respect to the total number of 
hydrogen nuclei ought to be less than 3$\times$ 10$^{-7}$.
This is in fact the 
case up to $>$ 9 $\times$ 10$^4$ yr. While we do not have such as 
accurate age for the Class I protostar, it is likely that its age does not exceed 
this value. 
In addition, note that beams of 10$\arcsec$--16$\arcsec$, as those
used to observe the present NO emission imply severe beam dilution 
for an emitting size of 1$\arcsec$ (typical of hot-corinos such
as the SVS13-A one; see De Simone et al. 17; Lef\`evre et al. 2017).
In particular, the filling factor is equal 
to 4 $\times$ 10$^{-3}$ (16$\arcsec$) and 
10$^{-2}$ (10$\arcsec$). 

In conclusion, 
the present dataset confirm that high-sensitivity observations 
allow one to reveal molecules (such as NO), whose emission can 
challenge the models so far used for protostellar systems.
Indeed high-angular interferometric
observations of NO emission are clearly  
needed 
to provide a comprehensive picture of the SVS13-A system. 

\subsection{Chemical modelling of L1157-B1}

UCLCHEM has been used to model the molecular emission from this object, 
as well as other shocked spots in low mass outflow, in several studies 
(Viti et al. 2011; Holdship et al. 2016; G\'omez-Ruiz et al. 2016; 
Holdship et al. 2017). For L1157-B1 the best fit models have 
so far indicated a shock velocity ranging from 20 to 40 km s$^{-1}$ and 
a pre-shock density of at least 10$^4$ cm$^{-3}$. 

For this work, we used the grid of models already ran in 
Holdship et al. (2017). In summary the grid covers preshock densities 
conservatively ranging from 10$^3$ to 10$^5$ cm$^{-3}$, and shock velocities 
from 10 to 65 km s$^{-1}$ (see their Table 2). 
In fact, based on previous work, we shall exclude from 
our discussion all models with a pre-shock density of 10$^3$ cm$^{-3}$; 
this is further justified by our excitation analysis in the previous 
section which implies a gas density of at least 10$^4$ cm$^{-3}$ (we note that even a pre-shock density of
10$^3$ cm$^{-3}$ implies a post-shock density of $\sim$ 4$\times$10$^3$ cm$^{-3}$, still lower than the density inferred from our excitation analysis of 10$^4$ cm$^{-3}$). 
We then find that the models which achieve a high enough abundance 
of NO to match the observations are: (i) either those with a pre-shock 
density of 10$^5$ and 10$^6$ cm$^{-3}$ with, respectively, 
45 and 40 kms$^{-1}$ shock velocities, or (ii) models with 
a pre-shock density of 10$^4$ cm$^{-3}$ but with a shock velocity 
of at least 60 kms$^{-1}$. These two sets of models have in common the 
maximum temperature the gas attains 
during the passage of the shock. 
While we can not completely exclude a preshock density of 10$^4$ cm$^{-3}$, 
we note that such a low pre-shock density requires a very fast shock, which is inconsistent with previous findings. 
Moreover the observed abundance for this model is reached for an age larger 
than 1000 years. A pre-shock density of 10$^5$ cm$^{-3}$ with 
a velocity of 45 km s$^{-1}$ (see Fig. 10), on the other hand, leads 
to an abundances of $\sim$ 10$^{-6}$ after a few hundred years, 
consistent with the dynamical time-scales derived by Podio et al. (2015).
The increase in NO abundance at $\sim$ 150 years is due to the reaction of nitrogen with OH: the latter is enhanced with time
due to the reaction of free oxygen with H$_2$. At later times, NO is more efficiently destroyed by the reaction 
with free nitrogen, which becomes more abundant when NH$_3$ starts decreasing (Viti et al. 2011).   
We note that the NO abundance is maintained through the dissipation length (see Fig. 10) but that, unlike the low velocity shock  
model presented in Chen et al. (2014), NO does {\it not} trace O$_2$. 
Clearly there must be a parameter space that one must 
explore when performing shock modelling where NO may be a good tracer 
of molecular oxygen.
This model would also be consistent with the best matching models 
we found for the NH$_3$, H$_2$O (Viti et. al. 2011), 
H$_2$S (Holdship et al. 2016) and phosphorous-bearing 
species (Lefloch et al. 2016).  

\begin{figure}
\begin{center}
\includegraphics[angle=0,width=8.5cm]{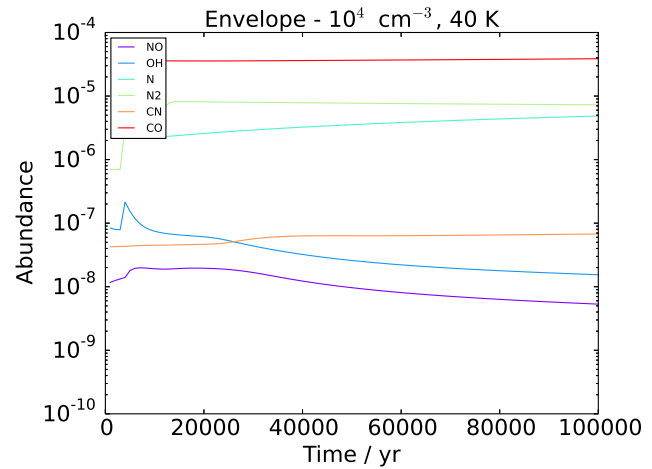}
\end{center}
\caption{Theoretical fractional abundances of 
NO, OH, and other selected species as a 
function of time, for Phase 2, for the envelope of SVS13-A for our best matching model,
assuming a density of 10$^4$ cm$^{-3}$ and a kinetic temperature 
of 40 K (see text).}
\end{figure}


\begin{figure}
\begin{center}
\includegraphics[angle=0,width=8cm]{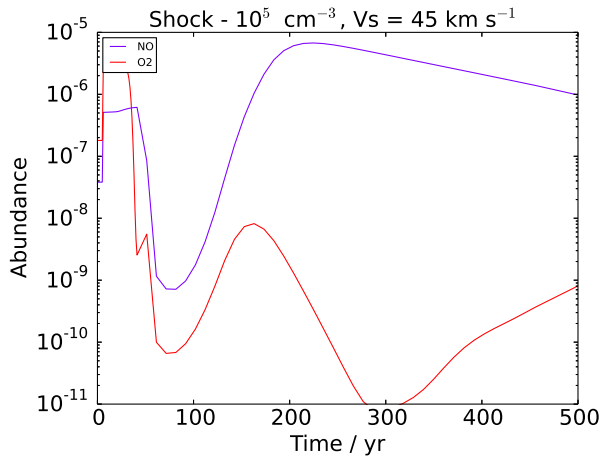}
\end{center}
\caption{Phase 2 Theoretical fractional abundances of NO (blue) and O$_2$ (red) 
for our best matching shock model up to the dissipation length, with a pre-shock density of 10$^5$ cm$^{-3}$ and a shock velocity of 45 km s$^{-1}$. The kinetic temperature of the pre-shock gas (Phase 1) is 10 K, while the post shock gas kinetic temperature is 80 K (Tafalla \& Bachiller 1995)}
\end{figure}

\section{Conclusions}

The high-sensitivity of the IRAM 30-m ASAI unbiased spectral survey
in the mm-window allows
us to detect towards the Class I object SVS13-A and the protostellar
outflow shock L1157-B1  
a large number of NO emission lines, namely the hyperfine
components of the $^2\Pi_{\rm 1/2}$
$J$ = 3/2 $\to$ 1/2 (at 151 GHz) 
and $^2\Pi_{\rm 1/2}$ $J$ = 5/2 $\to$ 3/2 (at 250 GHz) spectral
pattern. The main results of the analysis can be summarised as follows:

\begin{enumerate}

\item
Towards the SVS13-A system, narrow (1.5 km s$^{-1}$), 
optically thin ($\tau$ $\leq$ 0.2) lines peaking at the 
systemic velocity of +8.6 km s$^{-1}$ have been detected.
These findings, also supported by new CS(5--4),
C$^{18}$O, $^{13}$C$^{18}$O, C$^{17}$O, and $^{13}$C$^{17}$O lines
suggest the association of NO with
the molecular envelope with size of $\simeq$ 20$\arcsec$ 
(4700 AU) surrounding the SVS13-A protostar. 
The inferred excitation temperature $T_{\rm ex}$ is $\geq$ 4 K and the
total column density $N(\rm NO)$ $\leq$ 3 $\times$ 10$^{14}$ cm$^{-2}$;

\item
We observed NO emission towards L1157-B1:
the lines are broad (FWHM $\simeq$ 5 km s$^{-1}$) and 
blue-shifted by about 2 km s$^{-1}$, thus revealing for
the first time NO towards a shocked region driven by a low-mass protostar. 
The line profiles confirm that NO is associated with 
the extended (around 20$\arcsec$) cavity opened by the 
passage of the jet.
We derive $T_{\rm ex}$ $\simeq$ 5 K, and
$N(\rm NO)$ = 4--7 $\times$ 10$^{15}$ cm$^{-2}$.

\item
We derive a low NO fractional abundance of $\leq$ 3 $\times$ 10$^{-7}$
for the SVS13-A envelope, while a definite $X(NO)$ increase is
observed towards the L1157-B1 shock: $4-7$ $\times$ 10$^{-6}$.
To our knowledge, only few measurements of 
NO abundances towards low-mass protostellar regions have been so far reported, both around 10$^{-8}$ 
(extended PDR region around IRAS4A, Y{\i}ld{\i}z et al. 2013, and
starless cores, Akylmaz et al. 2007). The 
present measurements towards L1157-B1 support a huge
increase of the NO production in shocks.

\item
The public code UCLCHEM  
has been used to interpret the present NO
observations. We confirm 
that the abundance observed in SVS13-A can be attained by 
an envelope with
a gas density of 10$^5$ cm$^{-3}$ and a kinetic temperature of 40 K. 
The NO abundance in L1157-B1 
can be matched by a gas with a pre-shock density of 10$^5$ cm$^{-3}$ 
subjected to a shock with velocitiy of 40-45 kms$^{-1}$, in agreement 
with previous findings.

\end{enumerate}

\section*{Acknowledgements}

The authors are grateful 
to the IRAM staff for its help in the
calibration of the 30-m data. We also
thank the anonymous referee for instructive comments and suggestions.
The research leading to these results
has received funding from the European Commission Seventh Framework
Programme (FP/2007-2013) under grant agreement N. 283393 (RadioNet3).
This work was partly supported by
the Italian Ministero dell'Istruzione, Universit\`a e Ricerca through
the grant Progetti Premiali 2012 -- iALMA wich is also founding 
the EB PhD project.
This work was supported by the CNRS program ``Physique et Chimie du Milieu Interstellaire''
(PCMI) and by a grant from LabeX Osug@2020 (Investissements davenir - ANR10LABX56).
CF aknowledges support from the Italian Ministry of
Eductation, Universities and Research, project SIR (RBSI14ZRHR).
JH is funded by an STFC studentship (ST/M503873/1).
IJS acknowledges the financial support received from the STFC through an Ernest Rutherford Fellowship (proposal number ST/L004801).
BL and CCe acknowledge the
financial support from the French Space Agency CNES, and MT and RB
from Spanish MINECO (through project AYA2016-79006-P).

{}

\bsp

\label{lastpage}


\begin{thebibliography}{}

\bibitem[Akyilmaz et al.(1998)]{akyilmaz}
Akyilmaz, M., Flower, D.R., Hily-Blant P., Pineau des For$\hat {\rm e}$ts, G.,
\& Walmsley C.M. 2007, A\&A 462, 221   
\bibitem[Anglada et al.(2000)]{anglada}
Anglada, G., Rodr\'{\i}guez, L.F., Torrelles, J.M. 2000, ApJ 542, L123 
\bibitem[Asplund et al. (2009)]{asplund}
Asplund, M., Grevesse, N., Sauval, A. J., Scott, P, 2009, ARA\&A, 47, 481
\bibitem[Awad et al. (2016)]{awad16}
Awad, Z., Viti, S., Williams, D.A. 2016, ApJ 826, 207
\bibitem[Awad et al. (2010)]{awad10}
Awad, Z., Viti, S., Collings, M. P., Williams, D. A., 2010, MNRAS 407, 2511
\bibitem[Bachiller et al.(1998)]{bachiller98}
Bachiller, R., Guilloteau, S., Gueth, F., et al. 1998, A\&A 339, L49
\bibitem[Bachiller et al.(2001)]{bachiller01}
Bachiller, R., Per\'ez Guti\'errez, M., Kumar, M.S.N., \& Tafalla M. 2001, A\&A 372, 899
\bibitem[Blake et al. (1986)]{blake86}
Blake, G.A., Masson, C.R., Phillips, T.G., \& Sutton, E.C., 1986, ApJS 60, 357
\bibitem[Benedettini et al.(2013)]{benedettini13}
Benedettini, M., Viti, S., Codella, C., et al. 2013, MNRAS 436, 179
\bibitem[Busquet et al.(2014)]{busquet14}
Busquet, G., Lefloch, B., Benedettini, M., et al. 2014, A\&A 561, 120
\bibitem[Chen et al.(2009)]{chen}
Chen, X., Launhardt, R. \& Henning, Th. 2009, ApJ 691, 1729
\bibitem[Chen et al.(2014)]{chen14}
Chen, J-H, Goldsmith, P. F., Viti, S., Snell, R., et al. 2014, ApJ 793, 111
\bibitem[Chini et al.(1997)]{chini}
Chini, R., Reipurth, B., Sievers, A., et al. 1997, A\&A 325, 542
\bibitem[Codella et al.(1999)]{codella99}
Codella, C., Bachiller, R., \& Reipurth, B. 1999, A\&A 343, 585
\bibitem[Codella et al.(2010)]{codella2010}
Codella, C., Lefloch, B., Ceccarelli, C., et al. 2010, A\&A 518, L112
\bibitem[Codella et al.(2012)]{codella2012}
Codella, C., Ceccarelli, C., Bottinelli, S., et al. 2012, ApJ 744, 164
\bibitem[Codella et al.(2015)]{codella2015}
Codella, C., Fontani, F., Ceccarelli, C., et al. 2015, MNRAS 449, L11
\bibitem[Codella et al.(2016)]{codella2016}
Codella, C., Ceccarelli, C., Bianchi, E., et al. 2016, MNRAS 462, L75
\bibitem[Collings et al. (2004)]{coll04}
Collings, M.P., Anderson, M.A., Chen, R., et al. 2004, MNRAS 354, 1133
\bibitem[De Simone et al.(2017)]{desimone}
De Simone, M., Codella, C., Testi, L. et al. 2017, A\&A 599, 121
\bibitem[Flower et al. (2006)]{flower06}
Flower, D.R., Pineau des For$\hat {\rm e}$ts, G., Walmsley, C.M. 2006, A\&A 456, 215
\bibitem[Gerin et al. (1992)]{gerin92}
Gerin, M., Viala, Y., Pauzat, F., \& Ellinger, Y. 1992, A\&A 266, 463
\bibitem[Gerin et al. (1993)]{gerin93}
Gerin, M., Viala, Y., \& Casoli, F. 1993, A\&A 268, 212
\bibitem[Gomez-Ruiz et al.(2015)]{gomezruiz15}
G\'omez-Ruiz, A.I., Codella, C., Lefloch, B., et al. 2015, MNRAS 446, 3346
\bibitem[Gueth et al.(1996)]{gueth96}
Gueth, F., Guilloteau, S., \& Bachiller, R. 1996, A\&A 307, 891
\bibitem[Gueth et al.(1998)]{gueth98}
Gueth F., Guilloteau S., \& Bachiller R. 1998, A\&A 333, 287
\bibitem[Halfen et al. (2001)]{halfen01}
Halfen, D. T.; Apponi, A. J.; Ziurys, L. M., 2001, ApJ, 561, 244
\bibitem[Hily-blant et al. (2010)]{hily10}
Hily-Blant, P., Walmsley, C.M., Pineau des For$\hat {\rm e}$ts, G., \& Flower, D. 2010, A\&A 513, 41
\bibitem[Hirota et al.(2008)]{hirota}
Hirota, T., Bushimata, T., Choi, Y.K., et al. 2008, PASJ 60, 37
\bibitem[Holdship et al. (2017)]{hold17}
Holdship, J., Viti, S., Jim\'enez-Serra, I., \& Priestley, F., 2017, ApJ, submitted
\bibitem[Lique et al. (2009)]{lique}
Lique F., van der Tak F.F.S., Klos J., Balthuis J., \& Alexander M. 2009, A\&A 493, 557
\bibitem[Liszt \& Turner (1978)]{ll78}
Liszt, H.S., \& Turner B.E. 1978, ApJ, 224, 73
\bibitem[Lefevre et al.(2017)]{lefevre}
Lef\`evre C., Cabrit S., Maury A., et al. 2017, A\&A 640, L1
\bibitem[Lefloch et al.(1998)]{lefloch98}
Lefloch, B., Castets, A., Cernicharo, J., et al. 1998, A\&A, 334, 269
\bibitem[Lefloch et al.(2016)]{lefloch16}
Lefloch, B., Vastel, C., Viti, S., et al. 2016, MNRAS 462, 3937
\bibitem[Lefloch et al.(2017)]{lefloch17}
Lefloch, B., Ceccarelli, C.,  Codella, C., et al. 2017, MNRAS 469, L73 
\bibitem[Looney et al.(2000)]{looney}
Looney, L.W., Mundy, L.G., \& Welch, W.J. 2000, ApJ 529, 477
\bibitem[Looney et al.(2007)]{looney07}
Looney, L.W.; Tobin, J.J., \& Kwon, W. 2007, ApJ 2007, 670, L131 
\bibitem[Lopez-Sepulcre et al.(2015)]{lopez15}
L\'opez-Sepulcre, A., Jaber, A.A., Mendoza, E., et al. 2015, MNRAS 449, 2438
\bibitem[Martin et al. (2003)]{marin03}
Mart\'{\i}n, S., Mauersberger, R., Mart\'{\i}n-Pintado, J., Garc\'{\i}a-Burillo, S., \& Henkel, C., 2003, A\&A 411, 465
\bibitem[Mouschovias et al. (2006)]{mous06}
Mouschovias T. C., Tassis K., Kunz M. W., 2006, ApJ, 646, 1043
\bibitem[M\"uller et al.(2005)]{muller}
M\"uller, H.S.P., Schl\"oder, F., Stutzki, J., et al. 2005,
Journal of Molecular Structure, 742, 215
\bibitem[M\"uller et al.(2001)]{}
M\"uller, H.S.P., Thorwirth, S., Roth, D.A., et al. 2001, A\&A 370, L49
\bibitem[Nummelin et al. 2000]{nummelin00}
Nummelin, A., Bergman, P., Hjalmarson, \r{A}., et al. 2000, ApJS 128, 213
\bibitem[Podio et al.(2015)]{podio15}
Podio, L., Codella, C., Gueth, F. et al. 2015, A\&A 581, 85
\bibitem[Santangelo et al. (2015)]{santangelo}
Santangelo G., Codella C., Cabrit S., et al. 2015, A\&A 584, A126
\bibitem[Tafalla \& Bachiller(1995)]{teb} 
Tafalla, M., \& Bachiller, R. 1995, ApJ 443, L37
\bibitem[Tobin et al.(2010)]{tobin10}	
Tobin, J.J., Hartmann, L., Looney, L.W., Chiang, H.-F. 2010, ApJ 712, 1010
\bibitem[Tobin et al.(2016)]{tobin16}
Tobin, J.J., Looney, L.W., Li, Z.-Y., et al. 2016, ApJ 818, 73
\bibitem[Velilla Prieto et al. (2015)]{vell15}
Velilla Prieto, L., S\'anchez Contreras, C., Cernicharo, J., et al. 2015, A\&A 575, 84
\bibitem[Viti et al. (2004)]{viti04}
Viti, S., Collings, M.P., Dever, J.W., McCoustra, M.R.S., Williams, D.A., 2004, MNRAS, 354, 1141
\bibitem[Viti et al. (2011)]{viti11}
Viti, S., Jim\'enez-Serra, I., Yates, J.A., et al. 2011, ApJ, 740, 3 
\bibitem[Wilson et al.(1994)]{wilson94}
Wilson, T.L., \& Rood, R. 1994, ARA\&A 32, 191
\bibitem[Yildiz et al.(2014)]{yildiz}
Y{\i}ld{\i}z, U.A., Acharyya, K., Goldsmith, P.F., et al. 2013, A\&A 558, A58

\end{thebibliography}
\end{document}